\documentclass[a4paper,11pt]{article}
\pdfoutput=1 

\usepackage{jcappub} 

\usepackage[T1]{fontenc} 
\usepackage{float}
\usepackage{hyperref}
\usepackage{caption}
\usepackage{subcaption}
\usepackage{amsmath}
\usepackage[dvipsnames]{xcolor}

\title{Testing Regular Black Holes with X-ray data of GX~339--4}

\author[a,b]{Shafqat Riaz,}
\author[b]{Michail Kyriazis,}
\author[c,d,a]{Askar B. Abdikamalov,}
\author[a,c,1]{Cosimo Bambi, \note{Corresponding author.}}
\author[a]{and Swarnim Shashank }

\affiliation[a]{Center for Astronomy and Astrophysics, Center for Field Theory and Particle Physics,\\and Department of Physics,
Fudan University, Shanghai 200438, China}
\affiliation[b]{Theoretical Astrophysics, Eberhard-Karls Universit\"at T\"ubingen,\\D-72076 T\"ubingen, Germany}
\affiliation[c]{School of Natural Sciences and Humanities, New Uzbekistan University,\\Tashkent 100007, Uzbekistan}
\affiliation[d]{Ulugh Beg Astronomical Institute, Tashkent 100052, Uzbekistan}

\emailAdd{shafqat.riaz@uni-teubingen.de}
\emailAdd{michail.kyriazis@student.uni-tuebingen.de}
\emailAdd{a.abdikamalov@newuu.uz}
\emailAdd{bambi@fudan.edu.cn}
\emailAdd{swarnim@fudan.edu.cn}

\abstract{Regular black holes are singularity-free black hole spacetimes proposed to solve the problem of the presence of spacetime singularities that plagues the black holes of general relativity and most theories of gravity. In this work, we consider the regular black holes recently proposed by Mazza, Franzin \& Liberati and we extend previous studies to get a more stringent observational constraint on the regularization parameter $l$. We study simultaneous observations of \textit{NuSTAR} and \textit{Swift} of the Galactic black hole in GX~339--4 during its outburst in 2015. The quality of the \textit{NuSTAR} data is exceptionally good and the spectrum of the source presents both a strong thermal component and prominent relativistically blurred reflection features. This permits us to measure the regularization parameter $l$ from the simultaneous analysis of the thermal spectrum and the reflection features. From our analysis, we find the constraint $l/M < 0.44$ (90\% CL), which is stronger than previous constraints inferred with X-ray and gravitational wave data.}

\begin{document}
\maketitle
\flushbottom

\section{Introduction}\label{sec:intro}

One of the most outstanding and longstanding problems of general relativity is the presence of spacetime singularities in physically relevant solutions of the Einstein Equations. At a spacetime singularity, predictability is lost and standard physics breaks down. It is widely believed that the problem of spacetime singularities is a symptom of the breakdown of classical general relativity and it should be solved by quantum gravity. However, we do not have any mature theory of quantum gravity yet and we do not know the actual fate of these spacetime singularities.

In general relativity, the complete gravitational collapse of a body leads to the formation of spacetime singularities~\cite{Penrose:1964wq,Hawking:1970zqf}, which -- according to the Cosmic Censorship Conjecture~\cite{Penrose:1969pc} -- must be hidden behind an event horizon and the final product of the collapse should thus be a black hole. For a non-rotating (Schwarzschild) black hole, every particle crossing the black hole event horizon reaches the central singularity in a finite proper time. Since the spacetime is geodesically incomplete at the singularity, it turns out that we do not know what happens at the material swallowed by the black hole.

Since we do not have yet a robust theory of quantum gravity to solve the problem of the presence of spacetime singularities at the centers of black holes, we can try an alternative approach. We can construct singularity-free spacetimes and explore the implications of the resolution of singularities. Such a bottom-up strategy was discussed for the first time by James Bardeen at the end of the 1960s~\cite{bardeen68}, but only recently it has attracted the interest of a larger community and has become a hot topic; for a recent review on different attempts, see Ref.~\cite{book}. In the literature, we can find now a number of proposals for singularity-free black hole spacetimes~\cite{Dymnikova:1992ux,Ayon-Beato:1998hmi,Dymnikova:2003vt,Hayward:2005gi,Nicolini:2005vd,Ansoldi:2006vg,Spallucci:2008ez,Bambi:2013ufa,Bambi:2015zch,Frolov:2016pav,Bambi:2016wdn,Modesto:2021wdd}, singularity-free gravitational collapse models~\cite{Frolov:1981mz,Bambi:2013caa,Bambi:2013gva,Bambi:2016uda}, and even singularity-free cosmological models~\cite{Bojowald:2001xe,Ashtekar:2006wn,Ashtekar:2006rx,Cai:2011tc,Qiu:2011cy,Alexander:2014eva}.

The aim of the present work is to infer observational constraints on the possibility of the existence of such regular black holes, extending recent efforts to test the actual nature of astrophysical black holes~\cite{Bambi:2015kza,Yagi:2016jml}. We consider the regular black hole metric proposed by Mazza, Franzin \& Liberati in Ref.~\cite{Mazza:2021rgq}, which is the rotating version of the Simpson-Visser metric~\cite{Simpson:2018tsi,Simpson:2019cer,Lobo:2020ffi}. These regular black holes are characterized by the regularization parameter $l$, which has the dimensions of a length. For $l=0$, the spacetime reduces to the singular Kerr black hole solution of general relativity. For $l > 0$, the spacetime has no singularities and can describe either a regular black hole or a wormhole, depending on the values of the spin parameter $a_*$ and the regularization parameter $l$. In Ref.~\cite{Riaz:2022rlx} (hereafter Paper~I), we derived observational constraints on the regularization parameter $l$ from available X-ray and gravitational wave data. From the analysis of the reflection features of a \textit{NuSTAR} spectrum of the Galactic black hole in EXO~1846--031, we inferred the constraint $l/M < 0.49$ (90\% CL), where $M$ is the mass of the source and throughout this manuscript we use natural units $G_{\rm N} = c = 1$. From the LIGO-Virgo data of the gravitational wave event GW190707A, we found $l/M < 0.72$ (90\% CL). Here we want to extend the study in Paper~I and obtain a stronger constraint on the regularization parameter $l$. We report the analysis of a \textit{NuSTAR} spectrum (with a \textit{Swift} snapshot) of the Galactic black hole in GX~339--4 that simultaneously present strong reflection features and a prominent thermal component of the disk. Such a spectrum was already studied in Ref.~\cite{Tripathi:2020dni} and therefore we know that the quality of those \textit{NuSTAR} data is exceptionally good and that the simultaneous analysis of the reflection features and the thermal component has the capability to provide more stringent constraints than those possible from the sole analysis of the reflection features of a source~\cite{Tripathi:2020dni,Zhang:2021ymo,Tripathi:2021rqs}.

The manuscript is organized as follows. In Section~\ref{sec:metric}, we briefly review the regular black hole metric proposed by Mazza, Franzin \& Liberati in Ref.~\cite{Mazza:2021rgq}. In Section~\ref{sec:model}, we outline the construction of the lamppost model that we use in this work to fit the reflection features in the spectrum of GX~339--4. In Section~\ref{sec:data_analysis}, we present the reduction of the \textit{NuSTAR} and \textit{Swift} data. In Section~\ref{sec:sp_analysis}, we describe our spectral analysis. Discussion and conclusions are reported in Section~\ref{sec:discussion_conclusion}.

\section{Regular black holes}\label{sec:metric}

This section provides a succinct overview of the regular black hole metric employed in this study, as initially introduced in Ref.~\cite{Mazza:2021rgq}, for the reader's ease of comprehension. The metric is known as the rotating Simpson-Visser black hole, and it modifies the Kerr metric by introducing an additional parameter $l$. In Boyer-Lindquist coordinates  ($t, r, \theta, \phi$), the line element reads 
\begin{equation}
\label{line_element}
 {\rm d}s^2 = -\left(1 - \frac{2 M \sqrt{r^2 + l^2}}{\Sigma}\right){\rm d}t^2 + \frac{\Sigma}{\Delta}{\rm d}r^2 + \Sigma {\rm d}\theta^2 - \frac{4 M a \sin^2\theta \sqrt{r^2 + l^2}}{\Sigma}{\rm d}t{\rm d}\phi + \frac{A \sin^2\theta}{\Sigma}{\rm d}\phi^2 ,
\end{equation}
where
\begin{equation*}
\begin{gathered}
\Sigma = r^2 + l^2+ a^2 \cos^2\theta \, , \qquad
\Delta = r^2 + l^2 + a^2  - 2 M \sqrt{r^2 + l^2}  \,,
\\
A = \left(r^2+a^2 + l^2 \right)^2-\Delta a^2 \sin^2\theta \, .  
\end{gathered}
\end{equation*}
The variables $M$ and $a$ denote the mass of the central object and its spin angular momentum, respectively (the dimensionless spin parameter used in this manuscript is $a_* = a/M$). The aforementioned parameter denoted by $l$ serves as a regularization parameter that effectively excises the spacetime singularity, but only when $l > 0$. It is worth noting that $l$ possesses the length dimensions and can be interpreted as the deviation from the Kerr spacetime. The metric~\ref{line_element} is not related to the Kerr solution by any coordinate transformation; however, the Kerr solution can be obtained by setting the value of $l=0$. Furthermore, the metric remains invariant when subjected to the reflection symmetry (i.e., $r \rightarrow  -r$); this implies that the spacetime geometry can be perceived as comprising two indistinguishable segments connected at $r = 0$. The metric is regular everywhere when $l>0$ (as demonstrated in appendix A of Ref.~\cite{Mazza:2021rgq}), the singularity is excised, and now $r = 0$ is a regular surface of an oblate spheroid of size $l$, which an observer may cross. However, the metric posses coordinate singularities, which depend upon the value of $a$ and $l$ and can be obtained by setting $\Delta = 0$. The aforementioned singularities are associated with the spacetime horizons and are located at radial coordinates
\begin{equation}
r_{\rm H \pm} = \left( \left( M \pm \sqrt{M^2 - a^2} \right)^2 - l^2 \right)^{1/2} \, . 
\end{equation}
The outer (event) and inner horizons are represented by $r_{\rm H+}$ and $r_{\rm H-}$, respectively. 

Using the metric in Eq.~\ref{line_element} to analyze X-ray binary spectra is highly intriguing due to the potential of distinct values of $l$ and $a$ to indicate the existence of diverse categories of black hole mimicker spacetimes. The central object can exhibit six distinct morphologies based on the values of the variables $a$ and $l$. Interested readers may refer to Fig. 1 in Ref.~\cite{Mazza:2021rgq} for further details. Nonetheless, here we provide the criteria for $a$ and $l$ corresponding to distinct black hole mimickers. A wormhole that can be traversed is achieved under the conditions of either $a>M$ and $l > 0$ or $a<M$ and $l > M+\sqrt{M^2 - a^2}$. A null wormhole featuring an extremal event horizon is obtained when $a < M$ and $l = M + \sqrt{M^2-a^2}$. A regular black hole having one horizon per side is derived when $a<M$ and $M - \sqrt{M^2 - a^2} < l <M + \sqrt{M^2 - a^2}$. A regular black hole having an inner and outer horizon per side is attained by imposing $a < M$ and $l < M - \sqrt{M^2 - a^2}$. An extremal regular black hole possessing an extremal horizon per side is obtained when $a = M$ and $l < M$. Lastly, a regular black hole with one horizon per side and a null throat is obtained when $a< M$ and $l = M - \sqrt{M^2 - a^2}$.

In Paper~I, we implemented this metric in the reflection model ${\tt relxill\_nk}$~\cite{Bambi:2016sac,Abdikamalov:2019yrr,Abdikamalov:2020oci} (but only for the case of broken power-law emissivity profile) and in the thermal model ${\tt nkbb}$~\cite{Zhou:2019fcg} to analyze the X-ray data of accreting black holes. Hence, the development details of the aforementioned models are omitted, and they are directly implemented for the purpose of data analysis.

\section{\tt relxilllp\_nk}\label{sec:model}

As aforementioned, the metric was implemented in the model {\tt relxill\_nk}, but for arbitrary coronal geometry. An empirical radius-dependent disk's emissivity is commonly used to model the emissivity profile generated by a corona of unknown geometry: such an emissivity profile can take the form of a single power-law, a broken power-law, or a twice broken power-law. Here, we briefly describe the implementation of the metric within the framework of a specific geometry of corona, namely the lamppost corona~\cite{Wilkins:2012zm, Dauser:2013xv}, which is the most widely used coronal configuration; the resulting model is referred to  as {\tt relxilllp\_nk}~\cite{ Abdikamalov:2019yrr}.

The lamppost corona is a very simple model. It assumes that the corona is a point-like emitting source and can be completely described by its height, $h$, above the rotation axis of the black hole. The implementation of the metric is a two-step process. First, the transfer functions are computed for a grid of spin parameters ($a_*$), regularization parameters ($l/M$), and inclination angles ($i$), mapping the accretion disk to the observer's frame. A transfer function including all the relativistic effects of the spacetime (gravitational redshift, Doppler shift, and light bending) is tabulated for every point of the grid. For the spacetime in this study, the details of the transfer function calculations are presented in Paper~I, and therefore are not included here. As a second step, the disk's irradiation profile in the lamppost corona model is computed. This is done using a relativistic ray-tracing code, details of which are presented in Ref.~\cite{Abdikamalov:2019yrr}. To calculate the disk's incident intensity, photons are isotropically fired in the rest frame of the corona, and their trajectories are numerically computed until they hit the disk, i.e., $\theta = \pi/2$. By knowing the incident location of each photon, we can calculate the photon flux incident on the disk. Since the photons are  isotropically fired in equally spaced angles ($\delta$), the incident intensity can be calculated as 
\begin{equation}
    I_{\rm i} (r, h) = \frac{{\rm sin}(\delta)}{A(r, \Delta r) \gamma},  
\end{equation}
where $A (r, \Delta r)$ is the proper area of the ring on the disk where photons land, and $\gamma$ is the Lorentz factor of the disk, which includes the effect of the disk's rotation. Assuming the spectrum emitted from the corona of a power-law form with photon index $\Gamma$, the incident flux on the disk is
\begin{equation}
    F_{\rm i} (r, h) = I_{\rm i} g_{\rm lp}^{\Gamma}, 
\end{equation}
where $g_{\rm lp}$ is the photon redshift factor from the corona to the disk. This incident flux replaces the power-law emissivity profile in {\tt relxill\_nk}. The incident flux is computed for a grid of black hole spin parameters ($a_*$), corona heights ($h$), and regularization parameters ($l/M$) and stored in an additional FITS file. The grid for spin and regularization parameters is identical to that of the transfer function~\cite{Abdikamalov:2019yrr}.

\section{Data Reduction}\label{sec:data_analysis}

GX~339--4 is a low-mass Galactic X-ray binary system that experiences frequent outbursts every 1-3 years. It was first discovered by OSO-7 in 1973~\cite{Markert:1973}. Various X-ray missions have observed the object due to its frequent re-entry into the outburst phase. On March 4, 2015, \textit{Swift} monitoring identified robust thermal and power-law components in its spectrum. Subsequently, on March 11, 2015, the X-ray missions \textit{NuSTAR} (ID 80001015003) and \textit{Swift} (ID 00081429002) observed the source simultaneously~\cite{Parker:2016ltr}. The data reduction procedure was conducted according to the methodology described in Refs.~\cite{Parker:2016ltr, Tripathi:2020dni}.
             
First, we produced the cleaned event files from the XRT/\textit{Swift} data by applying the {\tt xrtpipeline} module of the \textit{Swift} XRT data analysis software (SWXRTDAS) v3.5.0. The software package is included in the Heasoft spectral analysis package v6.28 and CALDB version 20200724. The cleaned event files underwent additional processing and filtering through {\tt xselect} v2.4g. We extracted spectra from an annulus region surrounding the source to mitigate the pileup issue in the data. The annulus had an inner radius of 25 arcseconds and an outer radius of 45 arcseconds, which excluded the pileup region. The spectral redistribution matrices files (RMF) were used as given in CALDB. We generated the ancillary files using the module {\tt xrtmkarf}. Finally, the extracted data were binned to obtain a signal-to-noise ratio of 50. For the spectral analysis, we used the data in the 1-4~keV range.

The \textit{NuSTAR}'s raw data, observed for 30 ks, was processed into cleaned event files using the module {\tt nupipeline}. {\tt nupipeline} constitutes a component of the \textit{NuSTAR} data analysis software (NuSTARDAS) v2.0.0 and CALDB version 20200912. A circular region with a radius of 150 arcseconds, centered on the source, was selected as the source region. A 100-arcsecond region was chosen as the background, far from the source, to reduce the contamination of source photons. Finally, we extracted the spectra, RMF, and ARF files using {\tt nuproducts}. The FPMA and FPMB spectra were binned to obtain a signal-to-noise ratio of 50. We used the 3-40 keV band for subsequent analysis.

\begin{figure}[t]
\centering
\includegraphics[width=0.7\textwidth]{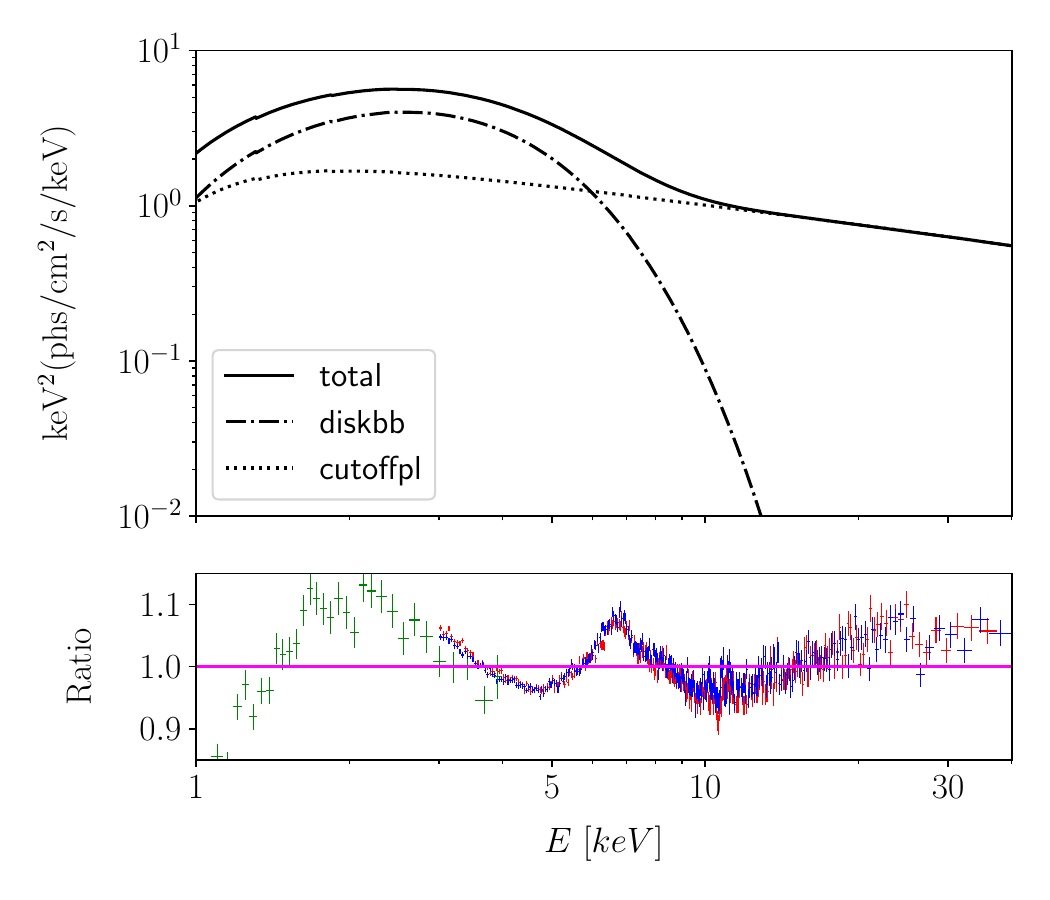}
\caption{Best-fit model (top panel) and data to best-fit model ratio (bottom panel) when we fit the \textit{NuSTAR} and \textit{Swift} data with {\tt tbabs*(diskbb + cutoffpl)} (Model~0). We use green crosses for XRT/\textit{Swift} data, red crosses for FPMA/\textit{NuSTAR} data, and blue crosses for \textit{NuSTAR}/FPMB data. \label{f-rm0}}
\end{figure}

\begin{figure}[t]
\centering
\includegraphics[width=1.0\textwidth]{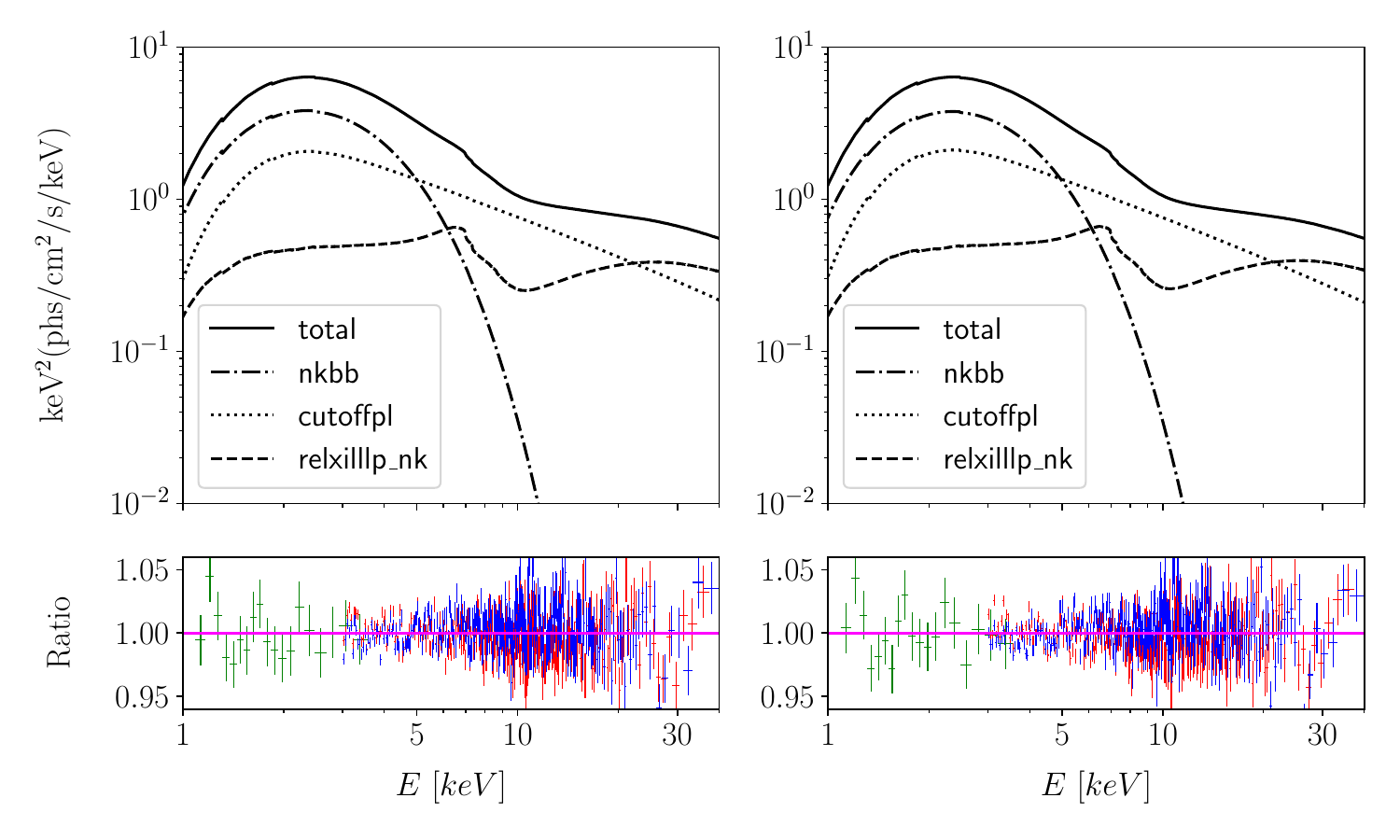}
\caption{Best-fit models (top panels) and data to best-fit model ratios (bottom panels) when we fit the \textit{NuSTAR} and \textit{Swift} data with {\tt tbabs*(cutoffpl + nkbb + relxilllp\_nk)} and we impose the Kerr metric (Model~1, left panels) and we leave the regularization parameter $l$ free in the fit (Model~2, right panels). We use green crosses for XRT/\textit{Swift} data, red crosses for FPMA/\textit{NuSTAR} data, and blue crosses for \textit{NuSTAR}/FPMB data. \label{f-mo-ra}}
\end{figure}

\section{Spectral Analysis}\label{sec:sp_analysis}

We follow the data analysis strategy presented in Refs.~\cite{Parker:2016ltr, Tripathi:2020dni}. Tab.~\ref{models} shows the list of models employed for our spectral analysis. Tabs. \ref{bft:m1-4}, \ref{bft:m5-7}, and \ref{bft:m8} present the best-fit values for Models~1--4, Models~5--7, and Model~8, respectively.

\subsection*{Model~0}
We begin by performing a joint fit of the \textit{NuSTAR}+\textit{Swift} data with an absorbed power-law plus a multi-color blackbody model (hereafter Model~0) as this combination frequently provides a reliable baseline for characterizing both high-energy Comptonized component from the corona and thermal component from the disk in X-ray spectra. In {\tt XSPEC}~\cite{xspec_software} syntax, the model reads
\begin{align}
    {\tt tbabs*(diskbb + cutoffpl)}. 
    \label{model0}
\end{align}
In~\ref{model0}, we model the X-ray spectrum with three components: (1) \texttt{tbabs}~\cite{Wilms:2000ez}, which accounts for the Galactic absorption, (2) \texttt{diskbb}~\cite{Mitsuda:1984nv, Makishima:1986ap}, a Newtonian model describing the disk’s thermal spectrum, and (3) \texttt{cutoffpl}, a power-law component with a high-energy cutoff that represents coronal emission. Fig.~\ref{f-rm0} shows the best-fit model (top panel) along with the ratio of data to this model (bottom panel). The residuals reveal a clear broadened iron line at around 7~keV and a pronounced Compton hump peaking between 20 and 30~keV in the \textit{NuSTAR}  data--features indicative of a strong reflection component.

\begin{table*}[tbh]
\renewcommand{\arraystretch}{1.3}
\centering
\begin{tabular}{c c c c c c c c}
\hline\hline
\vspace{-0.1cm}
\hspace{0.2cm} Models \hspace{0.2cm} & \hspace{0.2cm}  $l/M$ \hspace{0.2cm} & \hspace{0.2cm} emissivity~profile \hspace{0.2cm} & \hspace{0.2cm} $f_{\rm col}$ \hspace{0.2cm} & \hspace{0.2cm} Comptonization & \hspace{0.2cm} $E_{\rm cut}$ \\
\hline
Model~1 & $0^{*}$ & lamppost & 1.7$^{*}$ & {\tt cutoffpl} & free   \\
Model~2 & free & lamppost & 1.7$^{*}$  & {\tt  cutoffpl} & free \\
Model~3 & free & lamppost & 1.5$^{*}$  & {\tt  cutoffpl} & free \\
Model~4 & free & lamppost & 1.9$^{*}$  & {\tt  cutoffpl} & free\\
Model~5 & free & broken power-law & 1.7$^{*}$ & {\tt cutoffpl}& free \\
Model~6 & free & lamppost & 1.7$^{*}$ & {\tt nthComp} & $2\, kT_{\rm e}$ \\
Model~7 & free & lamppost & 1.7$^{*}$ & {\tt nthComp} & $3\, kT_{\rm e}$ \\
Model~8 & free & lamppost & 1.7$^{*}$ & {\tt compPS} & $2\, kT_{\rm e}$ \\
\hline\hline
\end{tabular} 
\vspace{0.2cm}
\caption{Summary of the differences in Models~1--8. We show (from left to right) the choices of the regularization parameter of the spacetime ($l/M$), the emissivity profiles of the disk in {\tt relxill$\_$nk}, the color-correction parameter ($f_{\rm col}$) in {\tt nkbb}, the Comptonization model for the coronal spectrum, and the high-energy cutoff of the coronal spectrum ($E_{\rm cut}$). $^*$ indicates parameters that are fixed at their specified values during the fitting process. \label{models} }  
\end{table*}

\subsection*{Model~1}
Next,  we improve Model~0 by introducing a relativistic reflection component {\tt relxilllp\_nk}~\cite{Bambi:2016sac,Abdikamalov:2019yrr,Abdikamalov:2020oci}, and replacing the Newtonian thermal component (\texttt{diskbb}) with the relativistic thermal model in non-Kerr spacetime (\texttt{nkbb})~\cite{Zhou:2019fcg}. With our goal in mind of constraining the regularization parameter, both \texttt{relxilllp\_nk} and \texttt{nkbb} adopt the spacetime described in Eq.~\ref{line_element}, linking their spin ($a_*$) and regularization parameter ($l/M$). We initially fix $l/M = 0$, corresponding to the Kerr metric, allowing direct comparison of our analysis with results reported in the literature. We set the color-correction factor in \texttt{nkbb} to $f_{\rm col} = 1.7$. This factor is a phenomenological parameter that compensates for non-thermal effects--primarily electron-photon scattering--in the disk’s atmosphere, which can alter the apparent thermal spectrum. A value of $f_{\rm col}=1.7$ is conventionally used for a $10~M_{\odot}$ black hole accreting at~$10\%$ of the Eddington limit~\cite{Shimura:1995nu, Davis:2006bk}. This configuration is referred to as Model~1 throughout this manuscript. The complete model is now expressed as follows:
\begin{align}
    {\tt tbabs*(cutoffpl + nkbb + relxilllp\_nk)}.
    \label{fit:model}
\end{align}
\texttt{relxilllp\_nk} models the reflection spectrum of an accretion disk around a regular black hole, assuming illumination by a lamppost corona. We set its reflection fraction to $-1$ so that no direct coronal continuum is included in this component; instead, the continuum is described by \texttt{cutoffpl}. In \texttt{nkbb}, we treat the black hole mass ($M$) and distance ($D$) as free parameters during the fit. Although the standard continuum-fitting method typically relies on $M$, $D$, and observer’s viewing angle $(i)$ from independent observations, we adopt a different approach. Here we tie the inclination ($i$) and spin parameter ($a_*$) in \texttt{nkbb} to the corresponding parameters in \texttt{relxilllp\_nk}, and let $M$ and $D$ be inferred from the thermal spectrum~\cite{Tripathi:2020dni}. This slightly unconventional approach helps maintain consistency between the reflection and thermal components while still probing the effects of varying mass and distance. The left column of Fig.~\ref{f-mo-ra} presents the best-fit result of Model~1 in the top panel and the ratio of the data to the best-fit model in the bottom panel.

\subsection*{Model~2}
We repeat the analysis of Model~1, this time allowing $l/M$ to vary as a free parameter in the fit. We refer to this modified version as Model~2. We have determined constraints on the black hole spin ($a_*$) and the regularization parameter ($l/M$). A detailed discussion of these constraints is presented in the next section. The right column of Fig.~\ref{f-mo-ra} presents the best-fit result of Model~2 in the top panel and the corresponding ratio of the data to the best-fit model in the bottom panel.

\subsection*{Models~3 and 4}
To investigate the impact of the color-correction factor, we refit Model~2 by setting $f_{\rm col}$ to~1.5 and~1.9  in {\tt nkbb}. We designate these configurations as Models 3 and 4, respectively. These analyses evaluates the systematic effects of the color-correction parameter, particularly on $a_*$ and $l/M$.

\subsection*{Model~5} Analyses of Models~3 and~4 indicate that $f_{\rm col}$ value of~1.7, as used in Model 2, yields the best statistical fit compared to other values. Therefore, we will base the remaining models on Model~2. To examine how the disk’s emissivity profile affects parameter estimation--particularly for $a_*$ and $l/M$--we repeat the analysis of Model~2 with a broken power-law emissivity profile for the reflection component. We refer to this new configuration as Model~5.

\subsection*{Model~6} Thus far, we have described the continuum emission from the corona using the simple empirical model \texttt{cutoffpl}. To investigate the impact of assumptions on the hot corona continuum, we substitute \texttt{cutoffpl} with \texttt{nthComp} in Model~2. We refer to this modified configuration as Model~6. We express the complete model as follows:
\begin{align}
    {\tt tbabs*(nthComp + nkbb + relxilllp\_nk)}.
    \label{fit:model}
\end{align}
\texttt{nthComp} is a more sophisticated model than \texttt{cutoffpl}. \texttt{cutoffpl} is a phenomenological model describing a power-law component with photon index $\Gamma$ and a high-energy cutoff $E_{\rm cut}$. \texttt{nthComp} is a Comptonization model describing a power-law component with photon index $\Gamma$ and both low-energy and high-energy cutoffs, which are determined, respectively, by the seed photon temperature $(kT_{\rm bb})$ and the coronal electron temperature ($kT_{\rm e}$). In our analysis, we tie $kT_{\rm e}$ in \texttt{nthComp} with $E_{\rm cut}$ in \texttt{relxilllp\_nk}: $E_{\rm cut} = 2\,kT_{\rm e}$.

\subsection*{Model~7} In our previous models, we set the high-energy cutoff of the coronal spectrum illuminating the disk as $E_{\rm cut} = 2\,kT_{\rm e}$, which serves as an approximation. To quantify the systematic bias this assumption may introduce into the constraints on $a_*$ and $l/M$, we introduce Model 7, where we adopt $E_{\rm cut} = 3\,kT_{\rm e}$ keeping all other assumptions identical to Model 2.

\subsection*{Model~8} We further investigate the impact of assumptions on the hot corona continuum by substituting \texttt{cutoffpl} with \texttt{compPS}~\cite{Poutanen:1996nv} in Model 2, designating this configuration as Model~8. The complete model is formulated as follows:
\begin{align}
    {\tt tbabs*(compPS + nkbb + relxilllp\_nk)}.
    \label{fit:model}
\end{align}
Following Model~6, we assume that the high-energy cutoff of the radiation illuminating the disk is $E_{\rm cut} = 2\,kT_{\rm e}$.

The discussion of our fits is presented in the next section.  

\begin{table*}[h]
         \renewcommand\arraystretch{1.1}
	\centering
	\vspace{0.5cm}
	\begin{tabular}{lccccc}
		\hline\hline
		Model & Model~1 & Model~2 &  Model~3 & Model~4 \\
		\hline
		{\tt tbabs}   \\
		$ N_{\rm H}/10^{22}$~cm$^{-2}$ &  $0.880_{-0.070}^{+0.035}$   & $0.910_{-0.032}^{+0.024}$ & $0.903^{+0.023}_{-0.024}$ & $0.917^{+0.033}_{-0.028}$ \\
		\hline
		{\tt nkbb}  \\ 
        $M$ [$M_{\odot}$] & $11.0^{+2.4}_{-2.8}$ & $11.4^{+2.3}_{-2.4}$ & $10.4^{+1.8}_{-1.1}$ & $11.5^{+1.7}_{-1.7}$  \\
        $\dot{M}$ [$10^{18}$~g~s$^{-1}$] & $0.45^{+0.20}_{-0.08}$ & $0.62^{+0.40}_{-0.19}$ & $0.58^{+0.24}_{-0.18}$ & $0.66^{+0.23}_{-0.24}$\\
        $D$ [kpc] & $10.0^{+1.4}_{-1.6}$ & $9.6^{+1.5}_{-0.8}$ & $10.4^{+1.7}_{-1.7}$ & $10.6^{+1.2}_{-1.3}$ \\
        $f_{\rm col}$ & $1.7^{*}$ & $1.7^{*}$ & $1.5^{*}$ & $1.9^{*}$\\
        $l/M$ & $0^{*}$ & $\dagger$ & $\dagger$ & $\dagger$ \\
        $a_*$ & $\dagger$ & $\dagger$ & $\dagger$ & $\dagger$ \\
        $i$ [deg] &  $\dagger$ & $\dagger$ & $\dagger$ & $\dagger$\\ 
        \hline
        {\tt cutoffpl} \\
		$ \Gamma$ &   $2.24^{+0.11}_{-0.22}$   & $2.30_{-0.12}^{+0.20}$ & $2.22^{+0.25}_{-0.23}$ & $2.23^{+0.17}_{-0.13}$ \\ 
        $E_{\rm cut}$ [keV] & $440^{+20}_{-30}$ & $410^{+24}_{-30}$ & $490^{+30}_{-20}$ & $460^{+27}_{-34}$\\
             	\hline 
		{\tt relxill(lp)\_nk}  \\ 
		$q_{\rm in}$ & $-$ & $-$ & $-$ & $-$ \\
		$q_{\rm out}$ & $-$ & $-$ & $-$ & $-$\\
		$r_{\rm br}$ & $-$ & $-$ & $-$ & $-$\\
		$ h$ [$r_{\rm g}$] &  $7.8^{+1.6}_{-1.7}$ & $7.5^{+1.4}_{-1.1}$ & $8.5^{+2.0}_{-1.5}$ & $8.6^{+1.5}_{-1.5}$  \\ 
		$ a_* $&  $0.990_{-0.009}$ & $0.992_{-0.009}$  & $0.960_{-0.009}$ & $0.986_{-0.008}$\\ 
		$i$ [deg] &  $30.1^{+1.8}_{-3.2}$ & $30.1^{+1.7}_{-2.8}$ & $32.3^{+2.0}_{-2.4}$ & $31.8^{+1.9}_{-2.4}$\\ 
		$\Gamma$ & $\dagger$ & $\dagger$ & $\dagger$ & $\dagger$ \\
		$\log\xi_{\rm in}$ [$\rm erg \cdot \rm cm \cdot \rm s^{-1}$]&   $3.62^{+0.18}_{-0.15}$ & $3.69^{+0.23}_{-0.31}$ &   $3.72^{+0.22}_{-0.23}$ & $3.76^{+0.34}_{-0.43}$ \\ 
		$ A_{\rm Fe} $ &   $4.5^{+0.4}_{-0.8}$ &  $4.6^{+1.2}_{-1.1}$ & $4.7^{+0.8}_{-1.8}$ &  $4.1^{+1.3}_{-0.5}$\\ 
		$E_{\rm cut}$ [keV]  & $\dagger$ & $\dagger$ & $\dagger$ & $\dagger$ \\
	$l/M$ & $0^{*}$ & $0.40^{+0.04}$ & $0.43^{+0.04}$ & $0.44^{+0.06}$\\
		\hline
		{\tt cflux} [$\rm erg~cm^{-2}~s^{-1}$] \\
		$\rm log(F_{\rm cutoffpl})$ & $-8.022^{+0.018}_{-0.015}$ & $-8.011^{+0.011}_{-0.011}$ & $-8.011^{+0.011}_{-0.011}$ & $-8.011^{+0.010}_{-0.009}$ \\
		$\rm log(F_{\rm nkbb})$ & $-7.820^{+0.021}_{-0.074}$ & $-7.830^{+0.020}_{-0.025}$ & $-7.831^{+0.024}_{-0.022}$ & $-7.829^{+0.022}_{-0.023}$ \\
		$\rm log(F_{\rm relxilllp\_nk})$ & $-8.42^{+0.09}_{-0.08}$ & $-8.38^{+0.04}_{-0.04}$ & $-8.37^{+0.03}_{-0.05}$ & $-8.39^{+0.04}_{-0.04}$ \\
		\hline
		{\tt constant} \\ 
		FPMA & $1^*$ & $1^*$ & $1^*$ & $1^*$\\
		FPMB & $1.001_{-0.002}^{+0.002}$ & $1.0150_{-0.0014}^{+0.0014}$ & $1.0153_{-0.0016}^{+0.0015}$ & $1.0154_{-0.0015}^{+0.0015}$ \\
        $C_{\rm XRT}$ & $0.993^{+0.009}_{-0.009}$ & $0.992^{+0.008}_{-0.008}$ & $0.995^{+0.007}_{-0.007}$ & $0.995^{+0.008}_{-0.008}$ \\
		\hline
		$\chi^2/\nu $ & $\quad 785.44/597 \quad$ & $\quad 781.85/596 \quad$ & $\quad 840.65/596$  & $\quad 850.65/596$ \\ 
		& $=1.31564$ & $=1.31183$ & $=1.41048$ & $=1.42726$\\
		\hline\hline
	\end{tabular}
	\vspace{0.2cm}
	\caption{Summary of the best-fit parameters for Models~1-4. The reported uncertainties correspond to the 90\% confidence level for one relevant parameter ($\Delta\chi^2=2.71$). $^*$ means the value of the parameter is frozen during the fit. $\dagger$ means the parameter is linked to the same parameter in another component. When there is no lower/upper uncertainty, the boundary of the range in which the parameter is allowed to vary is within the 90\% confidence limit. The flux of each model component is computed using the {\tt cflux} over the $0.5-50$ keV energy range.  \label{bft:m1-4}}
\end{table*}


\begin{table*}[h]
         \renewcommand\arraystretch{1.1}
	\centering
	\vspace{0.5cm}
	\begin{tabular}{lccc}
		\hline\hline
		Model & Model~5 & Model~6 & Model~7  \\
		\hline
		{\tt tbabs}   \\
		$ N_{\rm H}/10^{22}$~cm$^{-2}$ &  $0.915^{+0.020}_{-0.070}$   & $0.910_{-0.030}^{+0.030}$ & $0.912_{-0.027}^{+0.026}$   \\
		\hline
		{\tt nkbb}  \\ 
        $M$ [$M_{\odot}$] & $11.7^{+2.4}_{-2.7}$ & $11.2^{+2.2}_{-2.5}$ &  $11.2^{+2.1}_{-2.6}$   \\
        $\dot{M}$ [$10^{18}$~g~s$^{-1}$] & $0.68^{+0.22}_{-0.23}$ & $0.54^{+0.35}_{-0.28}$ & $0.57^{+0.30}_{-0.30}$  \\
        $D$ [kpc] & $11.2^{+1.5}_{-1.8}$ & $10.2^{+1.7}_{-1.7}$ &  $10.0^{+1.6}_{-1.1}$  \\
        $f_{\rm col}$ & $1.7^{*}$ & $1.7^{*}$ & $1.7^{*}$  \\
        $l/M$ & $\dagger$ & $\dagger$ & $\dagger$ \\
        $a_*$ & $\dagger$ & $\dagger$ & $\dagger$ \\
         $i$ [deg] &  $\dagger$ & $\dagger$ & $\dagger$ \\ 
        \hline
        {\tt cutoffpl} & \\
		$ \Gamma$ &   $2.21^{+0.12}_{-0.21}$   & $-$ & $-$  \\ 
        $E_{\rm cut}$ [keV] & $450^{+20}_{-20}$ & $-$ & $-$ \\
        \hline
		{\tt nthComp} \\
		$\Gamma$ &  $-$ & $2.17_{-0.12}^{+0.20}$ & $2.15^{+0.25}_{-0.23}$ \\
		$kT_{\rm e}$ [keV] & $-$ & $190^{+20}_{-35}$ &  $150^{+25}_{-30}$\\
    		$kT_{\rm bb}$ [keV] & $-$ & $0.34^{+0.13}_{-0.14}$ & $0.35^{+0.15}_{-0.15}$\\
		\hline
		{\tt relxill(lp)\_nk}  \\ 
		$q_{\rm in}$ & $5.8^{+1.6}_{-1.3}$ & $-$ & $-$  \\
		$q_{\rm out}$ & $2.80^{+0.10}_{-0.04}$ & $-$ & $-$ \\
		$r_{\rm br}$ & $5.4^{+0.4}_{-0.3}$ & $-$ & $-$ \\
		$ h$ [$r_{\rm g}$] &  $-$ & $7.9^{+1.2}_{-1.6}$ & $8.0^{+1.3}_{-1.7}$ \\ 
		$ a_* $&  $0.988_{-0.008}$ & $0.986_{-0.030}$ & $0.984_{-0.008}$  \\ 
		$i$ [deg]&  $30.2^{+1.7}_{-2.4}$ & $32.8^{+2.4}_{-3.3}$ & $31.6^{+1.5}_{-2.5}$  \\ 
		$\Gamma$ & $\dagger$ & $\dagger$ & $\dagger$  \\
		$\log\xi_{\rm in}$ [$\rm erg \cdot \rm cm \cdot \rm s^{-1}$]&   $3.64^{+0.23}_{-0.22}$ & $3.59^{+0.14}_{-0.32}$ & $3.55^{+0.16}_{-0.27}$ \\ 
		$ A_{\rm Fe} $&   $4.3^{+1.4}_{-0.9}$ &  $4.1^{+1.1}_{-0.4}$ & $4.2^{+1.3}_{-0.5}$ \\ 
		$E_{\rm cut}$ [keV]  & $\dagger$ & $2\, {T_{\rm e}}$ & $3 \, {T_{\rm e}}$ \\
		$l/M$ & $0.43^{+0.06}$ & $0.40^{+0.04}$ & $0.40^{+0.04}$  \\	
		\hline
		{\tt constant} \\ 
		FPMA & $1^*$ & $1^*$ & $1^*$  \\
		FPMB & $1.0152_{-0.0017}^{+0.0017}$ & $1.0151_{-0.0013}^{+0.0013}$ & $1.0152_{-0.0014}^{+0.0015}$  \\
        $C_{\rm XRT}$ & $0.993^{+0.013}_{-0.013}$ & $0.993^{+0.007}_{-0.007}$ & $0.994^{+0.007}_{-0.008}$  \\
		\hline
		$\chi^2/\nu $ & $\quad 815.94/594 \quad$ & $\quad 779.58/595 $ & $\quad 780.10/595$   \\ 
		& $=1.37363$ & $=1.31022$ & $=1.31109$ \\
		\hline\hline
	\end{tabular}
	\vspace{0.2cm}
	\caption{Summary of the best-fit parameters for Models~5-7. The reported uncertainties correspond to the 90\% confidence level for one relevant parameter ($\Delta\chi^2=2.71$). $^*$ means the value of the parameter is frozen during the fit. $\dagger$ means the parameter is linked to the same parameter in another component. When there is no lower/upper uncertainty, the boundary of the range in which the parameter is allowed to vary is within the 90\% confidence limit. \label{bft:m5-7}}
\end{table*}


\begin{table*}[bh]
         \renewcommand\arraystretch{1.1}
	\centering
	\vspace{0.5cm}
	\begin{tabular}{lcccc}
		\hline\hline
		Model & Model~8  \\
		\hline
		{\tt tbabs}   \\
		$ N_{\rm H}/10^{22}$~cm$^{-2}$ &  $0.915_{-0.025}^{+0.030}$   \\
		\hline
		{\tt nkbb}  \\ 
        $M$ [$M_{\odot}$] & $10.0^{+1.5}_{-1.5}$   \\
        $\dot{M}$ [$10^{18}$~g~s$^{-1}$] & $0.55^{+0.45}_{-0.30}$  \\
        $D$ [kpc] & $10.3^{+1.3}_{-1.2}$ \\
        $f_{\rm col}$  & $1.7^{*}$  \\
        $l/M$ &  $\dagger$ \\
        $a_*$ & $\dagger$\\
        $i$ [deg] &  $\dagger$  \\ 
                \hline
		{\tt compPS} \\
		$T_{\rm e}$ [keV] & $240^{+20}_{-35}$ \\
		$p$ &  $4.9$ \\
		$g_{\rm min}$ & $0.73$ \\
		$g_{\rm max}$ & $10$ \\	
		$T_{\rm bb}$ [keV] & $9.97^{+8}_{-8}$\\
		$\tau_{\rm y}$ & $<1.8$ \\
		\hline
		{\tt relxill(lp)\_nk}  \\ 
		$q_{\rm in}$  & $-$  \\
		$q_{\rm out}$ &  $-$ \\
		$r_{\rm br}$ &  $-$ \\
		$ h$ [$r_{\rm g}$] & $8.6^{+3.2}_{-2.6}$   \\ 
		$ a_* $ & $0.965_{-0.028}$  \\ 
		$i$ [deg]&  $35.8^{+2.8}_{-4.3}$  \\ 
		$\Gamma$ & $2.050^{+0.023}_{-0.016}$   \\
		$\log\xi_{\rm in}$ [$\rm erg \cdot \rm cm \cdot \rm s^{-1}$]&   $4.02^{+0.24}_{-0.24}$  \\ 
		$ A_{\rm Fe} $&    $3.4^{+1.1}_{-0.4}$  \\ 
		$E_{\rm cut}$ [keV]  & $2 \, {T_{\rm e}}$ \\
		$l/M$ & $0.55^{+0.32}$  \\	
		\hline
		{\tt constant} \\ 
		FPMA & $1^*$  \\
		FPMB &  $1.0151_{-0.0013}^{+0.0013}$  \\
        $C_{\rm XRT}$ & $0.993^{+0.007}_{-0.007}$   \\
		\hline
		$\chi^2/\nu $ & $\quad 926.85/591$   \\ 
		&  $=1.56827$ \\
		\hline\hline
	\end{tabular}
	\vspace{0.2cm}
	\caption{Summary of the best-fit parameters for Model~8. The reported uncertainties correspond to the 90\% confidence level for one relevant parameter ($\Delta\chi^2=2.71$). $^*$ means the value of the parameter is frozen during the fit. $\dagger$ means the parameter is linked to the same parameter in another component. When there is no lower/upper uncertainty, the boundary of the range in which the parameter is allowed to vary is within the 90\% confidence limit. In the case of $\tau_{\rm y}$ in {\tt compPS}, we report the 90\% confidence level upper limit. \label{bft:m8}}
\end{table*}

\section{Discussion and Conclusions}
\label{sec:discussion_conclusion}

\begin{figure}[t]
\centering
\includegraphics[width=0.70\textwidth]{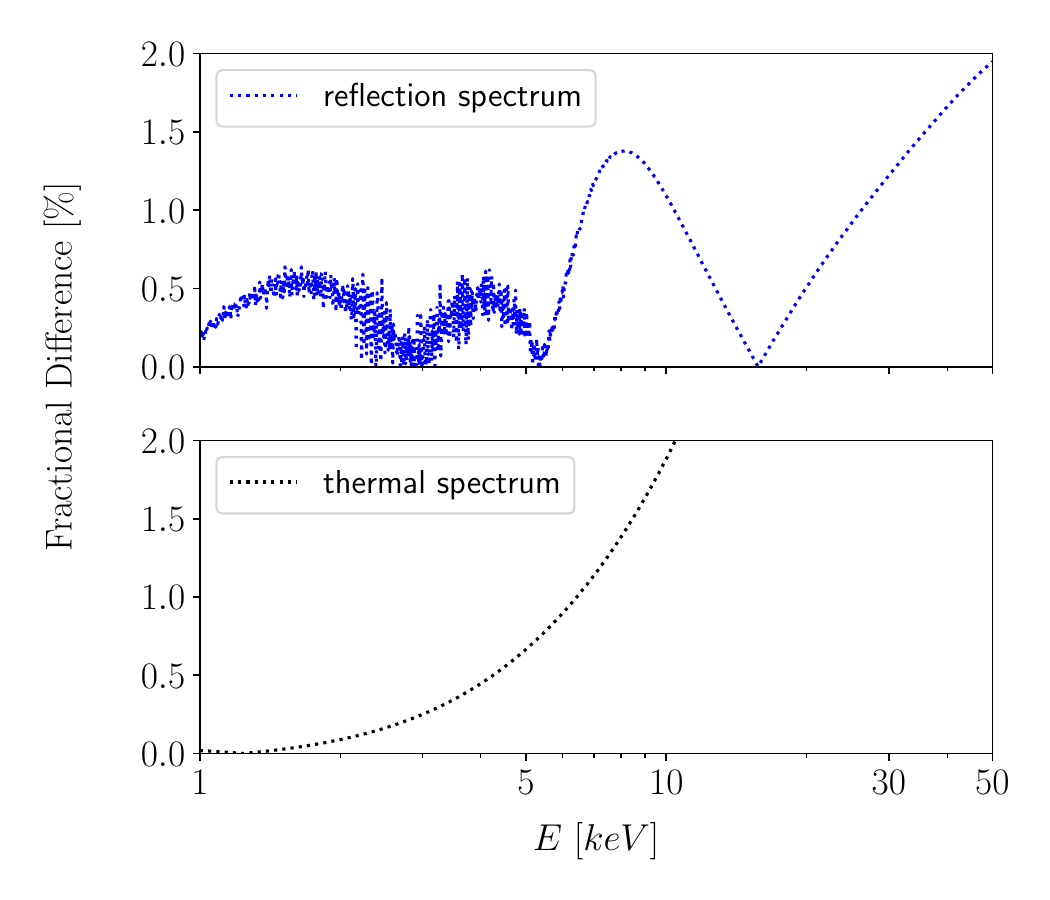}
\vspace{-0.2cm}
\caption{Comparison of theoretical spectra (top panel: reflection only; bottom panel: thermal only) between best-fit models for $l = 0$ (Model~1) and $l\neq 0$ (Model~2). The y-axis represents fractional difference, defined as $|F_{l\neq 0} - F_{l=0}|/F_{l=0}$, where $F_{l\neq 0}$ and $F_{l=0}$ denote theoretical spectral fluxes for $l\neq 0$ and $l= 0$, respectively. Key differences between the models are most pronounced around the iron line and Compton hump. While the thermal spectrum exhibits greater difference above 5 keV, its overall contribution to the total flux decreases at higher energies.  Note that all spectra were normalized at 0.5 keV.} \label{spectral_comparison}
\end{figure}

\begin{figure}[b]
\centering
\includegraphics[width=0.65\textwidth]{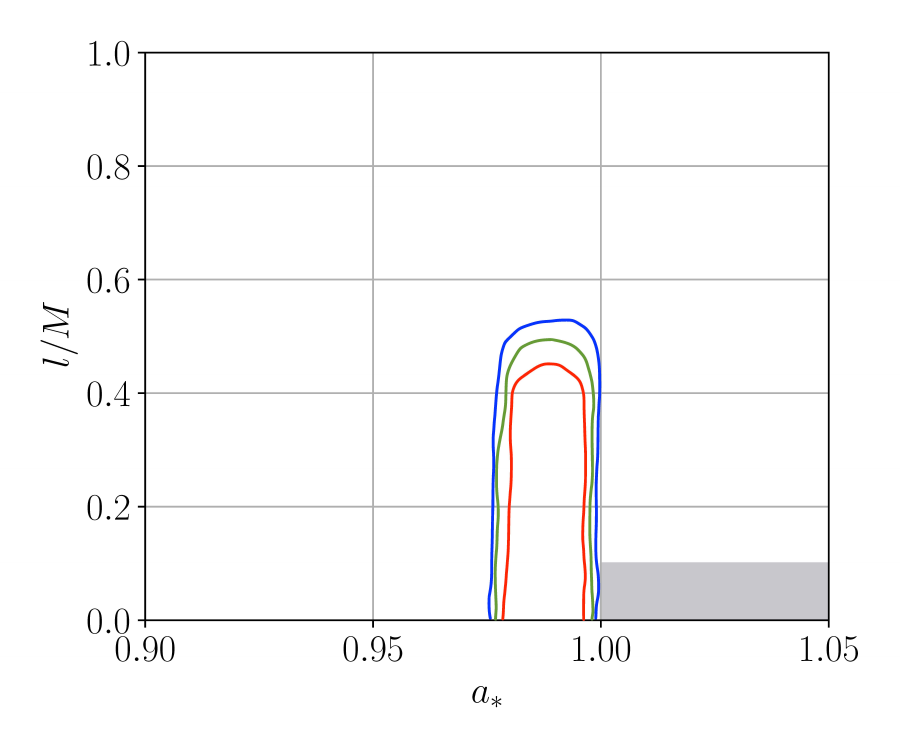}
\caption{Model~2 constraints on spin parameter $a_*$ and the regularization parameter $l/M$ from the analysis of \textit{NuSTAR} and  \textit{Swift} data of GX~339--4. The red, green, and blue curves represent the 68\%, 90\%, and 99\% confidence level contours for two relevant parameters, respectively. The gray region in the parameter space is excluded in our analysis in order to avoid the vicinity of spacetimes with a naked singularity. \label{contour}}
\end{figure}

We fit the data from the simultaneous observations with \textit{NuSTAR} and \textit{Swift} of the source GX~339--4 to test the Kerr hypothesis. The physical picture of the system is that of a black hole surrounded by a cold, geometrically thin, and optically thick accretion disk. There is also a compact and hot corona along the black hole spin axis. The thermal spectrum of the accretion disk is a multi-temperature blackbody spectrum and is described by the {\tt nkbb} model in our fits. Thermal photons from the disk can inverse Compton scatter off free electron in the corona. In our fits, the spectrum of the Comptonized photons is described by {\tt cutoffpl} ({\tt nthComp} in Models~6 and 7 and {\tt compPS} in Model~8). A fraction of the Comptonized photons illuminate the disk and produce the reflection spectrum, which is described by {\tt relxill\_nk}. The compact corona along the black hole spin axis can be the base of a jet and we use {\tt relxilllp\_nk}, which is the lamppost setup in the {\tt relxill\_nk} package. The lamppost setup is certainly quite a crude approximation, but it can fit the data quite well.

We consider the Kerr and the non-Kerr models (Model~1 and Models~2--8, respectively). The regular black hole metric is assumed as the background spacetime in the non-Kerr model~\cite{Mazza:2021rgq, Riaz:2022rlx}, and a constraint is derived for the spin ($a_*$) and regularization parameter ($l/M$). The quality of the fit for Models~1 and 2, the Kerr and the non-Kerr models with the same astrophysical assumptions, is good as $\chi^{2}_{\rm red}$ is close to 1 (see Tab.~\ref{bft:m1-4}), and there are no significant unresolved features in the ratio plots (see bottom panels of Fig.~\ref{f-mo-ra}). The non-Kerr fit exhibits a slightly lower $\chi^2_{\rm red}$ compared to the Kerr case, due to the inclusion of an additional parameter in the former fit (see Tab.~\ref{bft:m1-4}). Fig.~\ref{spectral_comparison} illustrates the spectral differences in reflection and thermal spectra by plotting the fractional difference between the best-fit Kerr (Model~1) and non-Kerr (Model~2) models. The primary difference between the reflection components of Model~1 and Model~2 is observed near the iron line (up to 1.5\%) and the Compton hump (up to 2\%). However, the Compton hump's contribution to the total flux diminishes towards higher energies. Similarly, the thermal components of Model~1 and Model~2 exhibit the most significant differences at higher energies, where their contribution to the total flux is minimal. Therefore, the iron line predominantly drives the constraint on the $a_*$ and $l/M$ parameters. 

In general, our results are consistent with those in Refs.~\cite{Parker:2016ltr, Tripathi:2020dni}; however, in contrast to Refs.~\cite{Parker:2016ltr, Tripathi:2020dni}, we report here the statistical uncertainties of 90\% confidence interval. There are some discrepancies in certain parameters, most importantly, the spin parameter. For instance 
\begin{equation}
a_* = 
    \begin{cases}
        0.95^{+0.02}_{-0.08}~(1 \sigma~\text{CL})  & \text{in  Ref.~\cite{Parker:2016ltr}} \\
        \\
        0.994^{+0.001}_{-0.002}~(1 \sigma~\text{CL}) & \text{in Ref.~\cite{Tripathi:2020dni}, B1 Model}\\
        \\
        0.990_{-0.009}~(90\%~\text{CL}) & \text{Model~1 in this work}\\
        \\
        0.992_{-0.009}~(90\%~\text{CL}) & \text{Model~2 in this work}\\
       	\\
	0.965_{-0.028}~(90\%~\text{CL})  & \text{Model~8 in this work}.
    \end{cases}
\end{equation}
The difference between our work and that of B1 fit in Ref.~\cite{Tripathi:2020dni} is not large. This is most likely owing to the fact that FITS files have the same precision and accuracy. When compared to Ref.~\cite{Parker:2016ltr}, however, the difference between the best-fit spin values is relatively considerable. 
The origin of this discrepancy may be attributed to the use of the older version of the {\tt relxilllp} model in Ref.~\cite{Parker:2016ltr} compared to our current FITS files, a slight difference in the accuracy of the FITS file, and a slight difference in the precision of the lamppost emissivity profile. Furthermore, we are comparing the best-fit results of the non-Kerr case with the Kerr, where the best-fit in the former is not exactly the Kerr, which introduces some further differences.

Regarding the Kerr hypothesis, we recover the Kerr metric as shown in Fig.~\ref{contour}. We present only the contour plot for Model~2. Our constraint on $l/M$ at the 90\% confidence level is $l/M < 0.44$ from Model~2, 6, and 7, which are the models that provide the best fits. In our analysis, we include only statistical uncertainty, which remains small because of the high quality of the data. Our measurement of the regularization parameter provides slightly tighter constraints than the measurement for the source EXO 1846--031 ($l/M < 0.49$ at 90\% confidence level) in Paper~I, which is currently the only known constraint from X-ray reflection spectroscopy. The stringent constraint on GX 339--4 may result from the combined effects of (i) the simultaneous fitting of \texttt{relxilllp\_nk} and \texttt{nkbb} for GX 339–4 and (ii) the higher-quality data available for this source. We note that in the fitting processes, we kept the black hole mass ($M$) and the distance ($D$) as free parameters. In contrast, these parameters are typically fixed at the values inferred from independent observations. The constraint on the regularization parameter ($l/M$) may be improved from independent estimates of $M$ and $D$, which are not impossible in the future.

Regarding other astrophysical parameters, our measured values for mass ($M$), distance ($D$), and mass accretion rate ($\dot{M}$) generally agree, within the stated uncertainties, with those reported in previous studies of this source~\cite{Parker:2016ltr, Tripathi:2020dni, Jiang:2019xqn}. Furthermore, our measurements of the disk's ionization parameter ($\log\xi$) demonstrate consistent values across our models when compared to prior determinations~\cite{Parker:2016ltr, Tripathi:2020dni, Jiang:2019xqn,Liu:2022bjr}. The iron abundance parameter ($A_{\rm Fe}$) presents a more complex situation. Our measurements indicate a super-solar value, consistent with findings in Refs.~\cite{Parker:2016ltr, Tripathi:2020dni}. While studies of the reflection features in the X-ray spectra of black hole X-ray binaries and active galactic nuclei often find values of $A_{\rm Fe}$ significantly higher than 1, the origin of such super-solar iron abundances is currently unknown. It may be the result of some deficiency in current reflection models (e.g., the actual electron densities may be higher than those in our models, two or more coronae may coexist at the same time and the X-ray radiation illuminating the disk and producing the reflection spectrum may not be the observed Comptonized continuum, the constant vertical electron density in our models may not be a good approximation, etc.) or even a true physical phenomenon (e.g., heavy elements may more easily move to the surface of the accretion disk); see, e.g., the discussion in Section~6 in \cite{Bambi:2020jpe} and references therein.

Modeling uncertainties can influence our constraints on  parameters. These uncertainties arise from simplifications in the models, which are often categorized into four classes: (i) simplifications in modeling the accretion disk, (ii) simplifications in modeling the hot corona, (iii) approximations in computing the local reflection spectrum, and (iv)  relativistic effects not considered properly~\cite{Taylor:2017jep, Riaz:2019bkv, Riaz:2019kat, Riaz:2020zqb, Riaz:2020svt, Riaz:2023xng, Bambi:2020jpe, Abdikamalov:2021rty, Abdikamalov:2021ues}.

Ref.~\cite{Tripathi:2020dni} studied the impact of modeling uncertainties on the test of the Kerr hypothesis using the Johannen metric~\cite{Johannsen:2012ng} with the same observations of the source GX~339--4. They studied the effect of different modeling assumptions, such as the choice of the emissivity profile,  the assumption on the color correction factor ($f_{\rm col}$) in {\tt nkbb}, the location of the inner edge of the disk ($R_{\rm in}$), electron density in the accretion disk ($n_{\rm e}$), and the presence of the cold material generating non-relativistic reflection spectrum. They concluded that the systematic uncertainties are under control and cannot significantly affect the measurements of key-parameters like the black hole spin and the deformation parameter. 

Following the methodology described in Ref.~\cite{Tripathi:2020dni}, we investigated the effect of $f_{\rm col}$ on parameter estimates, specifically $a_*$ and $l/M$, by fitting the data using two additional values ($f_{\rm col}$ = 1.5 and 1.9), corresponding to Models~3 and 4, respectively. The best-fit values are presented in Tab.~\ref{bft:m1-4}. Our results indicate that $\chi^2_{\rm red}$  values of Models~3 and 4 are marginally higher than those of Model~2; however, the best-fit values for $a_*$  and $l/M$ show no statistically significant deviation from those in model~2. The Kerr metric remains consistent with the data within the statistical uncertainties. Since the fit yields a slightly lower $\chi^2_{\rm red}$ value when $f_{\rm col} = 1.7$, we adopt this value in our remaining models.

In Models~1--4, we describe the corona in the reflection component as a lamppost geometry, likely an approximation of the actual coronal geometry in GX 339--4. The emissivity profile of the primary continuum is highly sensitive to the coronal geometry, which can significantly impact parameter estimates of black hole-disk properties, particularly in tests of the Kerr hypothesis~\cite{Riaz:2020svt}. To explore potential systematic biases when constraining $a_*$ vs $l/M$, we also fit the data using an emissivity profile modeled as a broken power-law (Model~5). The broken power-law emissivity profile is a phenomenological model defined as:
\begin{equation}
\epsilon \propto
    \begin{cases}
        1/r^{q_{\rm in}} &R_{\rm in}<r<R_{\rm br} \\
        \\
        1/r^{q_{\rm out}} &R_{\rm br}\le r<R_{\rm out},\\
    \end{cases}
    \label{emprof}
\end{equation}
where $R_{\rm br}$ is the breaking radius, $R_{\rm in}$ is the disk's inner radius, and $R_{\rm out}$ is the disk's outer radius.  In our analysis, the parameters $q_{\rm in}$, $q_{\rm out}$, and $R_{\rm br}$ are allowed to vary freely. The best-fit value for this model are presented in Tab.~\ref{bft:m5-7}. While the fit quality is slightly worse than that of Model~2, as indicated by a higher $\chi^2_{\rm red}$ value, the $l/M$ value for Model~5 deviates only slightly away from that of Model~2.  This deviation is within the given statistical uncertainties, and the Kerr metric is recovered, suggesting that the choice of emissivity profile does not introduce substantial systematic uncertainties into our measurement. Due to the improved fit quality obtained with the lamppost setup, we restricted our analysis to a lamppost-like emissivity profile for the corona in our subsequent models.

The lamppost-like corona setup naturally produces a very steep irradiation profile at the disk's inner region for a low-height source. Due to strong gravitational light bending at the disk's inner region, a significant fraction of the reflection component may return back to the accretion disk, namely the returning radiation, illuminating the disk's material producing so-called secondary reflection~\cite{Riaz:2023xng, Riaz:2020zqb, Dauser:2022zwc}. While the secondary reflection produces a distortion in the total reflection spectrum, we have not included the calculation of the secondary reflection in our model. This is because Ref.~\cite{Riaz:2023xng} suggested that the impact of returning radiation on our current capability of testing the Kerr black hole hypothesis is insignificant, particularly for a medium to large lamppost height.

In Model 6, we replaced {\tt cutoffpl} in Model~2 with {\tt nthComp} to describe the continuum emission spectrum from the corona. This was done to quantify the impact of the assumed continuum spectrum on the constraints on $a_*$ and $l/M$. The best-fit parameter values are reported in Tab.~\ref{bft:m5-7}. The $
\chi^2_{\rm red}$ values indicate that Models 2 and 6 provide a very similar fit.  Model~6 provides a slightly better fit, likely because it has one additional parameter. Regarding the constraints on $a_*$ and $l/M$ the two models yield very similar results, recovering the Kerr metric. Although {\tt cutoffpl} and {\tt nthComp} provide different spectra below 1 keV, we still obtain similar results. This is likely due to the absence of \textit{Swift} data below 1 keV.

In Model~6, we assumed that the high-energy cutoff ($E_{\rm cut}$) of the power-law spectrum from the corona illuminating the disk is related to the electron temperature ($kT_{\rm e}$) in {\tt nthComp} as $E_{\rm cut} = 2kT_{\rm e}$. However, this is an approximation and could introduce systematic uncertainties in our constraints. To quantify these potential uncertainties, we create Model~7, where we change $E_{\rm cut}$ from $2kT_{\rm e}$ to $3kT_{\rm e}$. The best-fit values are reported in Tab.~\ref{bft:m5-7}. We found that the measurements of Model~7 do not differ significantly from those of Model~2 and Model~6. This is likely because our analysis is limited to \textit{NuSTAR} data below 40~keV, where the effect of the $E_{\rm cut}$ assumption is minimal due to the coronal temperature significantly exceeds this energy range.

In Models~1--7, we employed either the {\tt cutoffpl} or {\tt nthComp} model to describe the Comptonized continuum from the corona. {\tt nthComp} is a more sophisticated model than the {\tt cutoffpl} yet rather a simple model. To explore the systematic bias associated with modelling the continuum from corona, we fit the data by replacing {\tt cutoffpl} in Model~2 with {\tt compPS}~\cite{Poutanen:1996nv} (Model~8).
{\tt compPS} is a more sophisticated model than the {\tt cutoffpl} and {\tt nthComp} and can potentially take into account several Comptonization geometries such as slab, sphere, hemisphere, and cylinder. For our analysis, a spherical geometry was utilized. Additionally, {\tt compPS} can include relativistic reflection from the cold disk material. However, in our analysis, we deactivated the reflection component in {\tt compPS} (by setting $rel_{\rm refl} = 0$) as it is already accounted by {\tt relxilllp\_nk}. Moreover, {\tt compPS} offers flexibility in defining the electron distribution function, allowing for Maxwellian, power-law, cutoff Maxwellian, or hybrid distributions, controlled by the parameter $p$, $g_{\rm min}$, and $g_{\rm max}$. We set these parameters to vary freely in the fit in order to account for these electron distribution plasma. The values of the best-fit parameters are reported in  Tab.~\ref{bft:m8}. We note that the best-fit value of the minimum Lorentz factor ($g_{\rm min}$) parameter in {\tt compPS} is smaller than 1 which means now the electron distribution is described by the Maxwellian distribution with parameter $T_{\rm e}$ and the non-thermal comptonization component is switch off. As observed, the {\tt compPS} model provides a poor fit in comparison to the other models and likely cannot be improved further probably due to its tendency to fit the thermal component near 1 keV. To avoid the tendency of the {\tt compPS} model to fit the thermal component, we switched off the {\tt nkbb}. The total model now becomes {\tt tbabs(compPS + relxilllp\_nk)}, and we repeated the entire fitting process. We got an even worse fit ($\chi_{\rm red}^2 \approx 1.62$), and we do not report the values of the best-fit parameters in this case. The reason for this bad fit is the unresolved residual close to 1.8~keV  suggesting that {\tt compPS} alone cannot fully account for the thermal components and a thermal component is needed to improve further the fit. We also fitted the data using other coronal geometries within the model, but those results are not included here due to similarly poor fit quality.

In conclusion, in Paper I, we put constraints on the regularization parameter of black hole mimickers spacetime using the X-ray data from the source EXO 1846\textendash{}031 and the gravitational wave data from black holes. The spacetime is phenomenological, and the regularization parameter removes the spacetime singularity. The precise nature of the regularization parameter is unknown at the moment; nonetheless, one can try to constrain it based on its macroscopic behavior; it controls the morphology of the compact object. Extending the work of Paper I, but only for the case of X-ray data, we derived constraints on the regularization parameter using the X-ray data from the simultaneous observation with \textit{NuSTAR} and \textit{Swift} of GX 339\textendash{}4. Thanks to the high-quality data as well as a broad ironline and a stronger thermal component, we obtained a slightly stronger constraint as compared to EXO 1846\textendash{}031. In order to drive a much stronger constraint, we need data from future X-ray missions, such as \textit{eXTP}~\cite{eXTP:2016rzs} or \textit{Athena}~\cite{Nandra:2013jka}, which promise to provide unprecedented high-quality data of accreting black hole systems.

\acknowledgments
This work was supported by the National Natural Science Foundation of China (NSFC), Grant No.~12250610185, 11973019, and 12261131497, the Natural Science Foundation of Shanghai, Grant No.~22ZR1403400, and the Shanghai Municipal Education Commission, Grant No.~2019-01-07-00-07-E00035.
S.R. acknowledges the support from Teach@Tuebingen fellowship.

\end{document}